\documentclass[pra,amsmath,amssymb,twocolumn]{revtex4}

\usepackage{amsmath,amssymb}
\usepackage[dvips]{graphicx}
\usepackage{feynmf}
\begin{document}

\title{Collective mode damping and viscosity in a 1D unitary Fermi gas}

\author{M. Punk}
\affiliation{Institute for Theoretical Physics, Universit\"at Innsbruck,
  Technikerstr.~ 25, A-6020 Innsbruck, Austria}
\author{W. Zwerger}
\affiliation{Physik-Department, Technische Universit\"at M\"unchen, James-Franck-Str. , D-85748 M\"unchen, Germany}

\date{\today}

\begin{abstract}
We calculate the damping of the Bogoliubov-Anderson mode 
in a one-dimensional two-component attractive Fermi gas for 
arbitrary coupling strength within a quantum hydrodynamic approach. 
Using the Bethe-Ansatz solution of the 1D BCS-BEC crossover 
problem, we derive analytic results for the viscosity covering the 
full range from a Luther-Emery liquid of weakly bound pairs
to a Lieb-Liniger gas of strongly bound bosonic dimers.
At the unitarity point, the system is a Tonks-Girardeau gas  
with a universal constant $\alpha_{\zeta}=0.38$
in the viscosity $\zeta=\alpha_{\zeta}\hbar n$ for $T=0$.  
For the trapped case, we calculate the Q-factor of the 
breathing mode and show that the damping provides 
a sensitive measure of temperature in 1D Fermi gases.
\end{abstract}

\pacs{03.75.Kk, 03.75.Ss, 74.20.Fg, 66.20.+d}

\maketitle


\section{Introduction}
\label{sec:Introduction}

In the past few years, ultracold atoms
have entered a new regime, where strong 
correlation effects appear even in 
extremely dilute gases. 
Prominent examples for this new area in atomic physics are the study 
of the crossover from a BCS-type superfluid of extended
Cooper pairs to a BEC of strongly bound molecules
\cite{Regal, Chin, Zwierlein} 
or the realization of a Tonks-Girardeau gas
of hard-core Bosons in one-dimensional atomic wires 
\cite{Paredes, Kinoshita}.  In the first case, the strong interaction 
regime is reached in a direct manner because the scattering length $a\/$ near a
Feshbach resonance becomes of the same order or even larger
than the average interparticle spacing $k_F^{-1}\/$. In the second case,
it is the squeezing of the kinetic energy in an optical lattice which 
enhances the role of interactions \cite{Zwerger}. 
 A unique role in the context of strongly interacting ultracold gases 
is played by the so-called unitary Fermi gas, where the dimensionless
interaction strength parameter $k_Fa$ is infinite. This problem was originally 
discussed in nuclear physics \cite{Baker, Heiselberg:2001}.
In its simplest form, it consists of a two-component Fermi gas with a zero
range attractive interaction which is just about to bind a state at  the two-particle
level. Such a situation is realizable with cold gases at a
Feshbach resonance, where the scattering length diverges \cite{OHara}.
Precisely at this point and for broad Feshbach 
resonances, where the range 
of the effective interaction is much smaller than the mean interparticle 
spacing, the full many-body problem 
has the bare Fermi-energy $\epsilon_F$ as the only energy scale.
As a result, the complete thermodynamics is a \emph{universal} function of the 
ratio $k_BT/\epsilon_F$ \cite{Ho}. While a quantitatively reliable description
of the many-body problem near a Feshbach resonance at finite temperature 
is still an open problem \cite{Bulgac, Burovski}, the situation
near zero temperature may be understood in a straightforward manner. 
Indeed,  at low temperatures, a two-component Fermi gas will be in a  
superfluid state, independent of the strength of the attractive interaction.
On quite general grounds therefore, the low lying excitations above the 
ground state are sound modes of the Bogoliubov-Anderson type,
which are the Goldstone modes of the broken gauge symmetry
in a neutral superfluid. In this regime, an effective low energy description 
is possible in terms of a quantum hydrodynamic approach \cite{LandauIX}.
For a Fermi gas with
a short range attractive interaction the associated effective field theory 
was recently discussed by Son and Wingate \cite{SW}. Starting from a Lagrangian 
formulation of the many-body problem, they 
realized that in the particular case of a unitary Fermi gas, there is  
an additional conformal symmetry with respect to arbitrary 
reparametrizations of the time.  
Remarkably, the effective field 
theory can be extended to nonequilibrium problems within the framework
of linear, irreversible thermodynamics. In particular, conformal invariance of the unitary 
Fermi gas applied to the dissipative part of the stress tensor requires
that two of the bulk viscosity coefficients vanish \cite{Son}.  As a result, no
entropy is generated in a uniform expansion. The extension of the effective
field theory to irreversible processes makes evident
that not only the thermodynamics but also dynamical properties like the 
kinetic coefficients are universal at the unitarity point, a fact,
first emphasized by Gelman, Shuryak and Zahed \cite{Gelman}. 
An example of particular interest is the shear viscosity $\eta$
which determines the damping of sound and collective oscillations
in trapped gases \cite{Massignan, Bruun}.  At unitarity, its dependence on 
density $n$ and temperature $T$ is fixed by dimensional arguments to be
$\eta=\hbar n\alpha(T/\mu)$, where $\mu$ is the chemical potential
and $\alpha(x)$ a dimensionless universal function \cite{Son}.  
At zero temperature, in particular, $\eta(T=0)=\alpha_{\eta}\hbar n$ 
is linear in the density with a universal coefficient $\alpha_{\eta}$.
Using a simple fluctuation-dissipation type argument in the
\emph{normal} phase, 
a lower bound of the form $\alpha_{\eta}\geq 1/6\pi$ has been derived 
by Gelman, Shuryak and Zahed \cite{Gelman}, in analogy to  
rigorous bounds for the ratio $\eta/\hbar s\geq 1/4\pi$ between
the viscosity $\eta\/$ and the entropy density $s\/$ in supersymmetric 
pure gauge Yang-Mills theories \cite{Policastro, Kovtun}. Based 
on these results, it has been speculated that ultracold atoms near 
a Feshbach resonance are a nearly perfect liquid \cite{Gelman}.

In the present work, we calculate the viscosity $\zeta\/$ of a strongly
interacting Fermi gas in the whole regime of coupling strengths
for the particular case of one dimension, where
an exact solution of the BCS-BEC crossover problem
has recently been given using the Bethe-Ansatz 
\cite{Tokatly, Fuchs, Recati}. It is shown that, at $T=0\/$,
the viscosity has the form $\zeta=\alpha_{\zeta}\hbar n\/$
with a coupling dependent constant $\alpha_{\zeta}\/$, which takes
the universal value $\alpha_{\zeta}=0.38\/$ at the unitarity 
point. At finite temperature, the sound damping does not have a
hydrodynamic form and increases like $\sqrt{T}\/$. We determine the
resulting damping of the breathing mode in a trapped gas and show
that its Q-factor provides a sensitive measure of temperature in
strongly interacting 1D gases.

\section{Sound Damping and Viscosity of a 1D Superfluid}
\label{sec:Viscosity}

\subsection{BCS-BEC Crossover in 1D}
Our calculations are based on an exactly solvable model 
of the BCS-BEC crossover in one dimension proposed 
by Fuchs et al. \cite{Fuchs, Recati} and by Tokatly \cite{Tokatly}.
The underlying microscopic Hamiltonian is that of the Gaudin-Yang model
\cite{GY} 
\begin{equation}
H=-\frac{\hbar^2}{2m}\sum_{i=1}^{N}\frac{\partial^2}{\partial x_i^2} +
g_1 \sum_{i<j}\delta(x_i-x_j)
\label{Ham}
\end{equation}
of a spin $1/2$ Fermi gas interacting via a short range
potential $g_1\delta(x)$.
Here $N$ is the total number of Fermions and $m$ their mass. At zero
temperature,  the model is characterized by a 
single dimensionless coupling constant $\gamma\equiv mg_1 / \hbar^2
n$, where $n\equiv N/L$ is the 1D density.  For attractive
interactions, the Hamiltonian (\ref{Ham}) describes a so-called Luther-Emery
liquid.  Its ground state at $\gamma \rightarrow 0^{-}$ is
a BCS-like state with Cooper pairs, whose size is much larger than the
average inter-particle spacing. With increasing 
magnitude of $\gamma$ one reaches the strong coupling regime
of  tightly bound molecules which behave like a hard
core Bose gas as $\gamma\to -\infty\/$. As shown by Girardeau, the hard core
Bose gas in one dimension is equivalent - for densities diagonal in real space -  to a gas
of {\it non}-interacting Fermions \cite{Girardeau}. Within a strictly 
one-dimensional model, the BEC-limit of strongly bound pairs is thus 
a Tonks-Girardeau gas. Now in practice, the atoms 
are trapped in a harmonic waveguide with radial frequency 
$\omega_{\perp}/2\pi$. The associated transverse oscillator length
$a_{\perp}\equiv\sqrt{\hbar/m\omega_{\perp}}$ then defines
an additional length, not present in the Gaudin-Yang model (\ref{Ham}). 
As shown by Bergeman {\it et al.} \cite{BMO}, the exact solution of the 
scattering problem for two particles in such a waveguide, interacting  
with a 3D pseudo-potential with scattering length $a\/$, always exhibits 
a two-body bound state,  \emph{whatever} the sign and magnitude of the 
scattering length $a$.  It appears at an energy $\hbar\omega_{\perp}-\tilde{\epsilon}_b$, 
which is below the continuum threshold at $\hbar\omega_{\perp}$ of
the transverse confining potential. Apart from this bound
state, all the scattering properties can be described by an effective
1D delta potential $g_1^{aa}\delta(x)$ for atom-atom interactions 
with strength \cite{Olshanii}
\begin{equation}
g_1^{aa}=2 \hbar \omega_{\perp}a \left( 1-Aa/a_{\perp}\right)^{-1}\, .
\label{g1olsh}
\end{equation}
As naively expected, an attractive 3D interaction $a<0$ implies a
negative value of $g_1^{aa}$. The associated binding energy
$\epsilon_b=m(g_1^{aa})^2/4\hbar^2$ in the 1D delta potential coincides 
with the exact value
$\tilde{\epsilon}_b$ in the weak confinement limit $\vert a\vert\ll a_{\perp}\/$.
Remarkably, the strength $g_1^{aa}\/$ of the 1D pseudo-potential remains
finite at a Feshbach resonance where $a=\pm \infty$. The
corresponding exact value of the binding energy is 
$\tilde{\epsilon}_b\simeq 0.6\hbar\omega_{\perp}$ \cite{BMO, Tokatly}.
Entering the positive side $a>0\/$ of the Feshbach resonance,  
the vanishing of the denominator in (\ref{g1olsh}) at
$a_{\perp}/a=A\simeq 1.0326$ \cite{sqrt2} leads to a 
confinement induced resonance (CIR), where $g_1^{aa}$
jumps from $-\infty$ to $+\infty$.  The exact bound state at  
this point has binding energy $\tilde{\epsilon}_b= 2\hbar\omega_{\perp}$ 
and a spatial extension along the $x\/$-axis, which is of the order of the
transverse oscillator length $a_{\perp}$.  
With decreasing values $a\lesssim a_{\perp}\/$ of the 
3D scattering length, $\tilde{\epsilon}_b\/$ increases monotonically
beyond $2\hbar\omega_{\perp}\/$ and finally approaches the 
standard 3D result $\tilde{\epsilon}_b\to\hbar^2/ma^2\/$ in the
weak confinement limit  $a\ll a_{\perp}\/$ \cite{BMO, Tokatly}. 
Since $\hbar\omega_{\perp}\gg \varepsilon_F\/$ in the limit of
a singly occupied transverse channel, the true bound state energy
$\tilde\epsilon_b$ is the largest energy scale in the problem in
the regime after the CIR where $g_1^{aa}>0\/$. In this regime,
the appropriate  degrees of freedom are no longer the single  
atoms but instead are strongly bound Fermion pairs,  
which are essentially unbreakable.  An exact solution
of the four-body problem in a quasi 1D geometry with tight harmonic
confinement shows, that these dimers have a \emph{repulsive}  
interaction in the regime beyond the CIR \cite{Mora}.  
The related constant $g_1^{dd}>0\/$ in the effective 
dimer-dimer interaction  $g_1^{dd}\delta(x)\/$ can be calculated 
as a function of the 3D scattering length \cite{Mora}.
It approaches  $g_1^{dd}\to 2\hbar\omega_{\perp}\cdot 0.6\, a\to 0\/$
in the weak confinement limit, where the dimer-dimer scattering length
$a^{dd}\approx 0.6\, a\/$ is identical with the one in free space \cite{Petrov}. 
Sufficiently far from the CIR, one thus recovers a weakly interacting gas 
of dimers. 

At the many-body level, the situation after the CIR is 
described by a Lieb-Liniger model \cite{LL} of repulsive Bosons.
Its dimensionless coupling constant 
$\gamma\equiv mg_1^{dd} / \hbar^2n$ is now positive and vanishes
in the weak confinement limit. It diverges at a value of the 
3D scattering length $a\/$ of order $a_{\perp}\/$. Now although 
the divergence of $g_1^{dd}\/$ does not exactly coincide with that 
of $g_1^{aa}\/$ at the CIR \cite{Mora},  the range of 
inverse dimensionless coupling constants where this mismatch 
appears is of order $1/\gamma\approx  na_{\perp}\/$ \cite{Recati}.
It is thus negligible
in the relevant low density limit $(na_\perp)^2\ll 1\/$.  Indeed,
at a fixed density $n\/$, the quasi 1D condition 
$\hbar \omega_{\perp}\gg\epsilon_F$ that only the lowest transverse 
mode is occupied is equivalent to  $(na_\perp)^2\ll 1$.    
In the limit $na_{\perp}\to 0\/$, there is thus a continuous evolution
from the Gaudin-Yang model of attractive Fermions to the Lieb-Liniger
model of repulsive Bosons which completely describes the BCS-BEC
crossover in one dimension \cite{Fuchs, Tokatly}. 
The associated spectrum of elementary excitations is straightforward
to understand: in the BCS limit $1/\gamma \to-\infty$,
the system consists of weakly bound Cooper pairs.  Their breaking 
is associated with a finite excitation gap and the corresponding
spectrum exhibits a relativistic dispersion relation 
$\varepsilon_s(k)=\sqrt{(\Delta/2 \hbar)^2+(v_s k)^2}$ 
similar to the standard quasiparticle spectrum of the BCS theory.
The associated energy gap $\Delta\/$ and the 
spin velocity $v_s>v_F$ increase monotonically with $1/\gamma\/$, 
both diverging in the strong coupling limit $1/\gamma=0\/$ at the CIR \cite{Fuchs}.
In addition, there are gapless density fluctuations describing
the Bogoliubov-Anderson mode of a neutral superfluid.  
These excitations exist for arbitrary coupling, both before and after
the CIR.  
Their spectrum is $\varepsilon(k)=v_c |k|$ at low momenta, with a (zero) 
sound velocity $v_c$ \cite{charge}, which monotonically decreases from the ideal 
Fermi gas value $v_c=v_F\/$ at $\gamma \rightarrow 0^{-}\/$ to the 
weak coupling BEC result $v_c=\sqrt{\gamma}v_F/\pi\/$ as 
$\gamma \rightarrow 0^{+}\/$ \cite{Fuchs}. 
 At the CIR, $1/\gamma=0$, the system is a 
Tonks-Girardeau gas \cite{Girardeau} of
tightly bound dimers.  The value $v_c(1/\gamma =0)=v_F/2\/$ 
simply reflects the fact that the unitary Fermi gas in 1D is a hard core 
Bose gas which - in turn - behaves like an ideal gas of spinless Fermions at half 
the original density. The universal parameter $\beta\/$, which 
follows from $v_c=v_F\sqrt{1+\beta}\/$ \cite{Ho}, thus has the exact value 
$\beta=-3/4\/$ in one dimension.  

\subsection{Quantum Hydrodynamic Theory}

In order to calculate the damping of long-wavelength phonons, we use 
a one-dimensional version of quantum hydrodynamics (QHD).  
Restricting the attention to the 
gapless Bogoliubov-Anderson mode in the superfluid regime, the QHD Hamiltonian has 
the same form on  both the Fermionic (before the CIR) and the Bosonic (after the CIR) side of the 1D BCS-BEC crossover. In a harmonic approximation, which is valid at low energies, the sound mode is described by the quadratic Hamiltonian
\begin{equation} 
H_0 = \: \frac{v_c}{2} \int_0^L dx \: \Big\{ \frac{\rho_0}{v_c} \; ( \partial_x\varphi)^2 + \frac{v_c}{\rho_0} \; \Pi^2 \Big\}
\label{H_0}
\end{equation}
where the conjugate fields $\varphi(x)$ and $\Pi(x)$ describe phase and density fluctuations 
respectively.  The only input parameters are the equilibrium mass-density $\rho_0=m n\/$  and the sound velocity $v_c\/$. From the Bethe Ansatz, the velocity $v_c\/$ is known as a function of the dimensionless inverse coupling constant $1/\gamma\/$, which ranges between $1/\gamma=-\infty\/$ in the BCS- and $1/\gamma=+\infty\/$ in the BEC-limit \cite{Fuchs}. To determine the damping due to the interaction of phonons, the energy functional of a one-dimensional quantum liquid needs to be 
expanded beyond quadratic order in the fields $\varphi(x)$ and $\Pi(x)$. It is a crucial advantage of the QHD approach, that the coefficients of the leading nonlinear terms are completely determined by thermodynamic quantities \cite{LandauIX}. Specifically, the lowest (third) order terms give rise to a contribution $H_{int}\/$ to the total Hamiltonian of the form
\begin{align}
\label{Hamiltonian}
H_{int} = & \: \frac{1}{6} \int_0^L dx \: \Big\{ (\partial_x \varphi) \Pi (\partial_x \varphi)+(\partial_x \varphi)^2 \Pi  \notag \\ + & \: \Pi (\partial_x \varphi)^2+\frac{d}{d\rho_0}\left(\frac{v_c^2}{\rho_0}\right) \Pi^3  \Big\}  
\end{align}
The quadratic Hamiltonian (\ref{H_0}) is diagonalized by the standard mode expansions
\begin{eqnarray*}
\varphi(x) &=& i \frac{\hbar}{m} \sqrt{\frac{\pi}{2}}\sqrt{\frac{v_c}{2 L \; v_F}} \; \sum_{q \neq 0} \frac{1}{\sqrt{|q|}} \left( b_{q} \; e^{i q x}-b^{\dag}_{q} \; e^{-i q x}\right) \\ 
\Pi(x) &=& m \sqrt{\frac{2}{\pi}} \sqrt{\frac{v_F}{2 L \; v_c}} \; \sum_{q \neq 0} \sqrt{|q|} \left( b_{q} \; e^{i q x}+b^{\dag}_{q} \; e^{-i q x}\right)  
\end{eqnarray*}
where $b_q^\dag$ and $b_q$ denote the usual bosonic creation and annihilation operators respectively and $v_F=\pi \hbar n /2 m$ is the Fermi velocity of the noninteracting gas. 
After inserting the mode expansions in (\ref{Hamiltonian}), we obtain
\begin{align}
H_{int}& = \sum_{q_1, q_2, q_3} \frac{1}{\sqrt{L}}\: V(q_1,q_2,q_3) \: \Big\{ \: b_{q_1} b_{q_2} b_{q_3} \delta(q_1+q_2+q_3) \notag \\ &+ b_{q_1} b_{q_2} b_{q_3}^\dag  \delta(q_1+q_2-q_3) + b_{q_1} b_{q_2}^\dag b_{q_3} \delta(q_1-q_2+q_3) \notag \\ &+ b_{q_1} b_{q_2}^\dag b_{q_3}^\dag \delta(q_1-q_2-q_3)+h.c. \: \Big\}
\end{align}
with the vertex
\begin{align}
V(q_1,q_2,q_3) & = \frac{\hbar^2}{6 m} \sqrt{\frac{\pi v_c |q_1 q_2 q_3|}{16 v_F}} \; \Big\{ \text{sgn}(q_1 q_3)+ \text{sgn}(q_1 q_2) \notag \\ & + \text{sgn}(q_2 q_3) + \frac{v_F^2}{v_c^2} \frac{d}{d v_F}\left(\frac{v_c^2}{v_F}\right) \Big\} 
\end{align}
The momentum dependence of the vertex makes perturbation theory applicable for long wavelength phonons. More precisely, it applies as long as the imaginary part of their energy is much smaller than the real part. As will be shown below, this requires the phonon wavelength $\lambda\/$ to be much larger than the interparticle spacing on the BCS side of the crossover, while in the BEC limit 
the more resctrictive condition $\lambda n\gg(1/\gamma)^{1/4}\gg 1\/$ is required. It should be remarked that an approach to evaluate the nonlinear terms in the 
Hamiltonian via Bosonization leads to additional coupling terms between phonons and spin excitations on the BCS side of the crossover which are not accounted for in the QHD approach. As will be shown in the Appendix, however, these terms do not contribute to the phonon damping at long wavelengths.

We evaluate the damping by calculating the imaginary part of the phonon self-energy $\Sigma(k,\omega)$ which is defined as the analytic continuation of the 
corresponding self-energy $\Sigma^{th}$ in the exact thermodynamic Green function
\[
\mathcal{G}(k,i \omega_n)=\frac{1}{i \omega_n-v_c |k|-\Sigma^{th}(k,i \omega_n)}\, . 
\]
Here $\omega_n=2 \pi n/\beta\/$ with $n\in \mathbb{Z}\/$ are the standard Bosonic Matsubara
frequencies. (We set $\hbar=k_B=1$ from now on, except in final results)
	
The main contribution to the damping rate comes from the three self-energy diagrams shown below, corresponding to spontaneous decay, absorption of a phonon and three wave annihilation. 
\begin{center}
	\begin{fmffile}{fmfdiag1}
	\begin{fmfchar*}(30,25)
 	 	\fmfleft{i1} \fmfright{o1}
  	\fmf{fermion,label=$k$}{i1,v1}
  	\fmf{fermion}{v2,o1}
  	\fmf{fermion,label=$k-q$,tension=0.2,left=0.5}{v1,v2}
  	\fmf{fermion,label=$q$,tension=0.2,left=-0.5}{v1,v2}
  	\fmfdot{v1,v2}
	\end{fmfchar*}
	\end{fmffile}
	
	\begin{fmffile}{fmfdiag2}
	\begin{fmfchar*}(30,25)
  	\fmfleft{i1} \fmfright{o1}
  	\fmf{fermion,label=$k$}{i1,v1}
  	\fmf{fermion}{v2,o1}
  	\fmf{fermion,label=$q-k$,tension=0.2,right=0.5}{v2,v1}
  	\fmf{fermion,label=$q$,tension=0.2,left=-0.5}{v1,v2}
  	\fmfdot{v1,v2}
	\end{fmfchar*}
	\end{fmffile}
	
	\begin{fmffile}{fmfdiag3}
	\begin{fmfchar*}(30,25)
  	\fmfleft{i1} \fmfright{o1}
  	\fmf{fermion,label=$k$}{i1,v1}
  	\fmf{fermion}{v2,o1}
  	\fmf{fermion,label=$-k-q$,tension=0.2,right=0.5}{v2,v1}
  	\fmf{fermion,label=$q$,tension=0.2,right=-0.5}{v2,v1}
  	\fmfdot{v1,v2}
	\end{fmfchar*}
	\end{fmffile}
\end{center}
Diagrams of this type have been considered before by Andreev \cite{Andreev} who studied the sound absorption in 1D Bose liquids for $T>0\/$ and by Samokhin \cite{Samokhin} in the context of the damping of zero sound in a 1D liquid of repulsive Fermions.

After taking the limit $L \rightarrow \infty$ and analytic continuation, the retarded self energy $\Sigma^R(k,\omega)$ is given by the sum of the three diagrams
\begin{equation}
\Sigma^R(k,\omega)=\Sigma^R_1(k,\omega)+2 \; \Sigma^R_2(k,\omega)+\Sigma^R_3(k,\omega)
\end{equation}
with
\begin{align*}
\Sigma^R_1(k,\omega) & = -18 \int \frac{dq}{2\pi}\frac{d \Omega}{2\pi} \; \coth(\frac{\beta \Omega}{2}) \; V^2(k,q,k-q) \\ & \times \Big\{ - G^R(q,\Omega+\omega) \; \text{Im}\: G^R(k-q,-\Omega) \\ & + G^R(k-q,\omega-\Omega) \; \text{Im}\: G^R(q,\Omega) \Big\}
\end{align*}
\begin{align*}
\Sigma^R_2(k,\omega) & = -18 \int \frac{dq}{2\pi}\frac{d \Omega}{2\pi} \; \coth(\frac{\beta \Omega}{2}) \; V^2(k,q,q-k) \\ & \times \Big\{G^R(q,\Omega+\omega) \; \text{Im}\: G^R(q-k,\Omega) \\ & + G^A(q-k,\Omega-\omega) \; \text{Im}\: G^R(q,\Omega) \Big\}
\end{align*}
\begin{align*}
\Sigma^R_3(k,\omega) & = -18 \int \frac{dq}{2\pi}\frac{d \Omega}{2\pi} \; \coth(\frac{\beta \Omega}{2}) \; V^2(k,q,-k-q) \\ & \times \Big\{-G^A(q,\Omega-\omega) \; \text{Im}\: G^R(-k-q,-\Omega) \\ & + G^A(-k-q,-\omega-\Omega) \; \text{Im}\: G^R(q,\Omega) \Big\}\, .
\end{align*}
Here $G^R$ and $G^A$ are the usual retarded and advanced Green functions respectively.
The combinatorial factor 18 arises from the three possible ways of choosing the creation/annihilation 
operators for the initial and final phonon and two possibilities of pairing the phonons in between.
As was already pointed out by Andreev and Samokhin, the fact that for linearly dispersing phonons in 1D, momentum- and energy conservation are simultaneously satisfied, requires to go beyond second order perturbation theory which would give an infinite damping rate. Following the approach of Andreev \cite{Andreev}, we calculate the self-energy by using  the fact that 
$\Sigma^R(k,\omega) \ll \varepsilon_k:=v_c|k|$ at long wavelengths. 
The precise condition on $k\/$ for which this holds, 
has to be determined afterwards and will be discussed below.  Since we are interested in the quasiparticle pole of $G^R(k,\omega)=1/(\omega-\varepsilon_k-\Sigma^R(k,\omega))$ we can use the approximation
\[
\omega=\varepsilon_k+\Sigma^R(k,\omega)\approx \varepsilon_k+\Sigma^R(k,\varepsilon_k)=:\varepsilon_k+\Sigma^R_k
\]    
leading to
\begin{equation}
G^R(k,\omega) \approx \frac{1}{\omega-\varepsilon_k-\Sigma^R_k}
\label{Greensfunc}
\end{equation}
The damping rate of phonons with wavevector $k>0\/$ can now be determined from the 
imaginary part $\Gamma_k:=\text{Im} \: \Sigma^R(k,\varepsilon_k) \equiv \text{Im} \: \Sigma^R_k$
of the (on-shell) self energy. 

\begin{figure}
	\begin{center}
		\includegraphics[width=0.9\columnwidth]{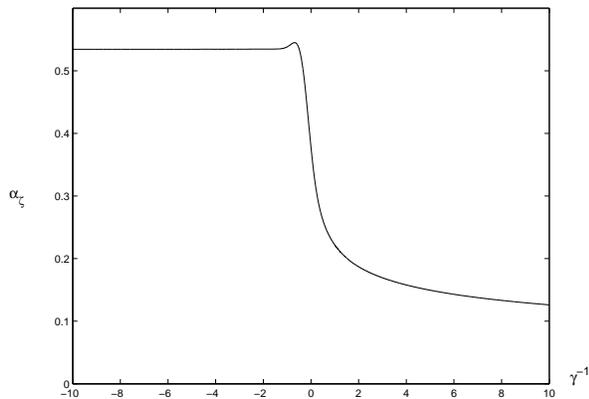}
	\end{center}
	\caption{Viscosity-parameter $\alpha_\zeta$ as a function of the inverse coupling constant $\gamma^{-1}$.}
	\label{fig:alpha0}
\end{figure}

Starting with the case of zero temperature $T=0\/$,  the only contribution to $\Sigma_k^R$ comes from spontaneus decay (first diagram). Applying the approximation (\ref{Greensfunc}) in the integrand and doing the $\Omega$- integration, one ends up with

\begin{align*}
\Sigma^R_k & = 9 \int \frac{dq}{2\pi}V^2(k,q,k-q) \: \Big\{G^R(q,\varepsilon_k-\varepsilon_{k-q}-\Sigma^R_{k-q}) \\ & + G^R(k-q,\varepsilon_k-\varepsilon_q-\Sigma^R_q) \Big\}\, .
\end{align*}      
The major contribution to the integral comes from $0<q<k$ where $\varepsilon_k-\varepsilon_{k-q}-\varepsilon_q=0$. Thus we arrive at the equation
\begin{align*}
\Sigma^R_k & = -\frac{\hbar^4 \pi v_c}{32 \: m^2 v_F}\left\{3+\frac{v_F^2}{v_c^2}\; \frac{d}{d v_F}\left(\frac{v_c^2}{v_F}\right)\right\}^2 \\ & \times \int_0^k \frac{dq}{2\pi}kq(k-q) \; \frac{1}{\Sigma^R_{k-q}+\Sigma^R_q}  
\end{align*}
for the retarded self energy. It is solved with a purely imaginary Ansatz
$\Sigma^R_q=-i \mu q^2$ where  
\begin{align*}
& \mu = \frac{\hbar^2}{4 m} \sqrt{\frac{\pi \:v_c}{2\: v_F}\; f_1(a=2)}\left\{3+\frac{v_F^2}{v_c^2}\; \frac{d}{d v_F}\left(\frac{v_c^2}{v_F}\right)\right\} \\ & f_1(a) := \frac{1}{2\pi} \int_0^1 dx \frac{x(1-x)}{(1-x)^a+x^a}
\end{align*}
At zero temperature, the resulting damping rate 
\begin{equation}
\Gamma_k^0=\frac{\hbar}{8 m} \sqrt{\frac{v_c}{v_F}\left(\frac{\pi}{4}-\frac{1}{2} \right) } \left\{ 3+ \frac{v_F^2}{v_c^2} \frac{d}{d v_F}\left(\frac{v_c^2}{v_F}\right)\right\} k^2
\label{damp_hom_T0}
\end{equation}
of the Bogoliubov-Anderson mode in one dimension is therefore quadratic in the wavevector. 
Formally, this is precisely the behaviour of a hydrodynamic mode. It allows to define a zero
temperature viscosity $\zeta\/$ by the relation
\begin{equation}
\label{viscosity}
\omega_k=v_c k-i \; \frac{\zeta}{2 m n} \;k^2
\end{equation}
which is completely analogous to sound damping in three
dimensions. In that case,  $\zeta\/$ is replaced by the combination $\zeta_2+4\eta/3\/$ 
involving one of the superfluid bulk viscosities $\zeta_2\/$ and the shear
viscosity $\eta\/$ \cite{Gelman, Forster} and one has $\zeta_2=0\/$ at unitarity  \cite{Son}. 
From the result (\ref{damp_hom_T0}), we see that the viscosity at zero temperature has the form
\begin{equation}
\zeta=\alpha_\zeta \hbar n
\label{visc}
\end{equation}
with a constant 
\begin{equation}
\label{alphazeta}
\alpha_\zeta := \frac{1}{4} \sqrt{\left(\frac{\pi}{4}-\frac{1}{2} \right) \frac{v_c}{v_F} } \left\{ 3+ \frac{v_F^2}{v_c^2} \; \frac{d}{d v_F}\left(\frac{v_c^2}{v_F}\right)\right\}\, .
\end{equation}
A plot of $\alpha_\zeta$ is given in figure \ref{fig:alpha0}, where the exact result for $v_c/v_F$ from the Bethe-Ansatz solution was used. Evidently, the dimensionless viscosity coefficient
$\alpha_\zeta$ depends on the inverse coupling constant $1/\gamma\/$ of the BCS-BEC 
crossover. It is thus in general dependent on the particle density $n\/$. At the unitarity point,
however, where $1/\gamma =0$, this dependence vanishes and $\alpha_\zeta$ takes
the universal value
\[
\alpha_\zeta(\gamma^{-1}=0)=\sqrt{\frac{\pi}{8}-\frac{1}{4}} \approx 0.38
\]
which is just $1/\sqrt{2}\/$ of the value $\alpha_{\zeta}(\gamma=0^{-})=0.54\/$ attained
in the weak coupling limit of the 1D noninteracting Fermi gas. 
Concerning the range of applicability of the perturbative calculation, it is obvious 
that the approximation leading to (\ref{Greensfunc}) is  satisfied for small wavenumbers 
$k\ll mnv_c/\zeta\/$. Based on the explicit result for $\zeta\/$ this condition  translates into
phonon wavelengths much larger than the mean interparticle spacing on the BCS side
of the crossover, including the unitarity point. On the BEC side, the ratio $v_c/\zeta\/$ vanishes
like $\sqrt{v_c}\sim\gamma^{1/4}\/$. The phonon wavelenghts have thus to obey the 
more restrictive condition  $\lambda n\gg(1/\gamma)^{1/4}\gg 1\/$ mentioned above. It is
interesting to note, that this condition is \emph{less} restrictive than the requirement
$\lambda\gg\xi_1\/$ for the validity of the linear dispersion relation $\omega_k=v_c |k|\/$,
where $\xi_1=n^{-1}(1/\gamma)^{1/2}\/$ is the 1D healing length.   

We now turn to the situation at finite temperature $T>0\/$, where
the long wavelength phonons, for which  $T \gg \varepsilon_k\/$,  behave classically. The thermal 
factor  $\coth(\beta \Omega/2)\/$ may therefore be replaced by its classical limit $2/(\beta \Omega)\/$.
At $T\ne 0\/$, the second diagram representing the absorption of another phonon also 
contributes to the damping.  An explicit calculation along the lines performed at $T=0\/ $ gives
a phonon damping rate at finite temperature of the form  
\begin{equation}
\Gamma_k^T=\frac{\hbar}{4 m} \sqrt{\frac{\pi}{2} \frac{k_B T}{\hbar v_F} f_2(3/2)} \left\{ 3+ \frac{v_F^2}{v_c^2} \frac{d}{d v_F}\left(\frac{v_c^2}{v_F}\right)\right\} k^{3/2}\, .
\label{dampT}
\end{equation}
Here $f_2(a=3/2)\approx 0.6221$ is a numerical coefficient defined by the integral
\[
f_2(a):= \int_0^1 \frac{dx}{2\pi}\frac{1}{(1-x)^a+x^a}+2 \int_1^\infty \frac{dx}{2\pi}\frac{1}{(x-1)^a+x^a}
\]
As already noted by Andreev \cite{Andreev}, the damping $\sim k^{3/2}$ for $T>0$ is not of the standard hydrodynamic form, in contrast to the behaviour at zero temperature. The quite 
different results can be understood from the fact that at any finite temperature, the 
quasi long-range superfluid order present at $T=0\/$ is destroyed by phase fluctuations
on a characteristic length scale $\xi_T:=\hbar v_c/k_B T\/$. Depending on the ratio $y=k\xi_T\/$ between
this length scale and the phonon wavelength, the behaviour is either essentially superfluid
for $y\gg 1\/$  or normal for $y\ll 1\/$. Similar to the formulation used in
dynamical scaling laws near critical points \cite{Ferrell, Halperin}, the   
crossover between the two different types of behaviour may be described by an
Ansatz of the form  
\begin{equation}
\Gamma_k=\frac{\hbar k^2}{2 m} \; \Phi (\xi_T k)\, .
\label{dampuniv}
\end{equation}
The associated crossover function has the limiting behaviour
\[
\Phi(y)=
\begin{cases}
		\ \alpha_\zeta & \text{for} \; y\rightarrow \infty \ (T \rightarrow 0) \\ \\
		\dfrac{3.70 \; \alpha_\zeta}{\sqrt{\:y}} & \text{for} \; y \ll 1 \ (T\gg \varepsilon_k)
\end{cases}
\]
with the parameter $\alpha_\zeta\/$ defined in Eq. (\ref{alphazeta}). It should be pointed out,
that the dependence of the 
damping rate on temperature is a simple power law $\sim T^{1/2}\/$ only to the extent that
the temperature dependence of the velocity $v_c\/$ itself can be neglected.  Moreover, 
note that for nonzero temperature, the damping (\ref{dampT}) remains finite in the BEC limit $1/\gamma \rightarrow \infty$ in contrast to the $T=0$ case.

\subsection{Harmonically trapped gas}

Finally we extend our results on damping in a homogeneous gas to the experimentally accessible 
case of a harmonically trapped system, using essentially a local density approximation. Our main interest is to calculate the damping of the so-called breathing modes, which have already been measured 
in 1D Bose gases in a regime near the Tonks-Girardeau limit \cite{Moritz:2003}.
Assuming a standard type of viscous damping in a classical fluid, the damping rate in a 
1D inhomogeneous case is given by \cite{LandauVI}
\begin{equation}
\Gamma= \left| \frac{\langle \dot{E}_{mech}\rangle_t}{2 \langle E \rangle_t} \right|= \left| \frac{\int dz \; \zeta(z) \langle (\partial_z v)^2 \rangle_t}{2 m \int dz \; n(z) \langle v^2 \rangle_t} \right|
\label{dampallg}
\end{equation} 
where $\langle . \rangle_t$ denotes the time average, $z$ is the spatial coordinate and the last equation holds for harmonically oscillating perturbations, where $\langle E \rangle_t = 2 \langle E_{kin} \rangle_t $.

\begin{figure}
	\begin{center}
		\includegraphics[width=0.78\columnwidth]{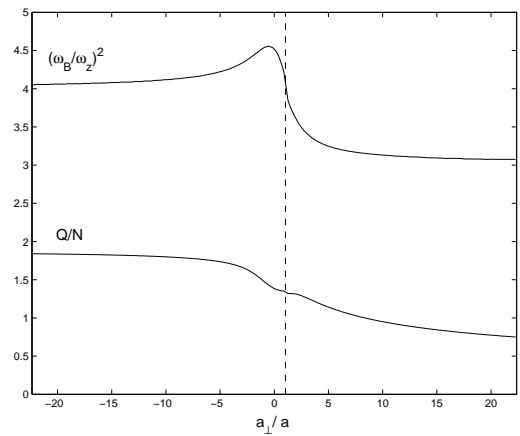}
	\end{center}
	\caption{Frequency $\omega_B$ and quality factor $Q=\omega_B/\Gamma_0$ of breathing modes at $T=0$ as a function of the 3D scattering length $a$. The dashed line indicates the CIR (unitarity point). The plot is for $\omega_{\perp}=5 N \omega_z\/$, where $\omega_\perp$ and $\omega_z$ are the radial and axial trapping frequency, N denotes the number of particles in the trap.}
	\label{fig:freq_Q_T0}
\end{figure}

Breathing modes in a harmonic trap are characterized by a velocity profile of the form $v(z,t)=\text{const.} \: z \: e^{-i \omega_B t}$.  Since the damping of the Bogoliubov-Anderson mode at 
zero temperature has precisely the form of a standard viscous fluid one obtains using Eq. (\ref{visc}) 
\begin{equation}
\Gamma_B^0 \; =\frac{\hbar}{2 m} \; \frac{\langle \alpha_\zeta \rangle}{\langle z^2 \rangle} 
\label{damp_breath_T0}
\end{equation}
where the brackets denote the spatial average defined by
\[
\langle f(z) \rangle = \dfrac{1}{N} \int dz \; n(z) f(z)
\]
and $N$ is the total number of particles. Since the constant $\alpha_{\zeta}\/$ depends on 
density except at the unitarity point, the damping also involves a spatial average
$\langle \alpha_\zeta \rangle\/$. 
A plot of the coupling dependent Q-factor $Q=\omega_B/\Gamma_B^0\/$ of the breathing mode 
at zero temperature is given in figure \ref{fig:freq_Q_T0} together with the ratio of its 
frequency in units of the axial trap frequency $\omega_z\/$.  The required density profiles were calculated numerically using a local density approximation and the exact results for the chemical potential from the Bethe-Ansatz solution. As shown by Menotti and Stringari \cite{Menotti}, the density profiles also determine the frequency from 
\[
\omega_B^2=-2\left(\frac{d\ln{\langle z^2\rangle}}{d\omega_z^2}\right)^{-1}
\]
The appearance of a maximum in the breathing frequency just before the
confinement induced resonance may qualitatively be understood from the
fact that around this point the nature of the pairing changes from
overlapping, correlated pairs to individual molecules.  Indeed, the size
of a molecule is of the order of the average inter-particle distance 
$n^{-1}\/$
for inverse coupling constants $1/\gamma\approx -0.5\/$ \cite{Fuchs}. 
Remarkably, this is also close to the point, where the effective dimer-dimer 
interaction $g_1^{dd}\/$ diverges \cite{Mora} and where the dimensionless viscosity
coefficient (\ref{alphazeta}) exhibits a small maximum as shown in Fig. \ref{fig:alpha0}.

For $T>0\/$ the situation is more complicated, because the $k\/$ -dependence of the damping rate in (\ref{dampT}) implies a non-hydrodynamic behaviour. 
In order to account for the specific $k\/$ -dependence in a classical calculation, we modify the stress-'tensor'  by introducing an effectively velocity dependent viscosity.
The form of $\zeta$ to be used in (\ref{dampallg}) which leads to the result (\ref{dampuniv}) for the damping in the homogeneous system is
\begin{equation}
\zeta=\hbar n \: \Phi \left( \xi_T \frac{\langle | \partial_z v | \rangle}{\langle | v | \rangle} \right)
\label{viscuniv}
\end{equation}  
with the function $\Phi$ as defined in (\ref{dampuniv}).

\begin{figure}
	\begin{center}
		\includegraphics[width=0.9\columnwidth]{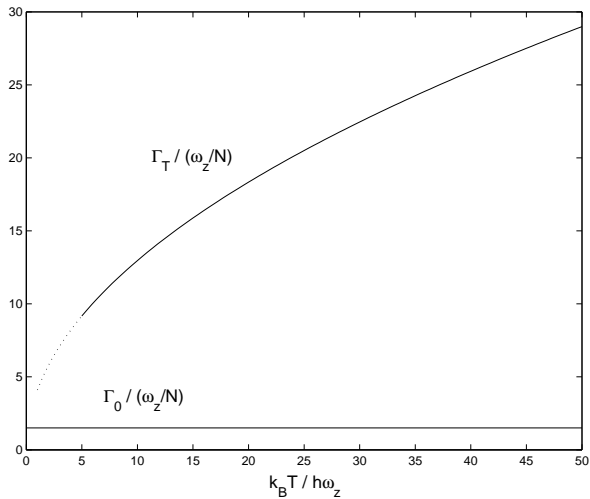}
	\end{center}
	\caption{Damping of breathing modes at the unitarity point as a function of $T$. The damping at $T=0$ is shown for comparison.}
	\label{fig:damp_T}
\end{figure}

Using (\ref{viscuniv}) for the inhomogeneous system, the result for the damping of breathing modes can be expressed as
\begin{equation}
\Gamma_{B}=\frac{\hbar}{2 m} \left\langle \Phi \left( \frac{\xi_T}{\langle |z| \rangle} \right) \right\rangle \frac{1}{\langle z^2 \rangle}
\end{equation}
In the particular case of the Tonks-Girardeau limit describing the unitary 1D Fermi gas at the CIR 
the damping of breathing modes at $T=0$ and at finite temperatures $T \gg \hbar \omega_z$
is given by  
\begin{eqnarray}
\Gamma_B^0 \; & \approx & 1.5 \; \frac{\omega_z}{N} \\
\Gamma_B^T \; & \approx & 4.1 \; \frac{\omega_z}{N} \sqrt{\frac{k_B T}{\hbar \omega_z}}
\end{eqnarray}
A plot is given in figure \ref{fig:damp_T}. Note that in the Tonks-Girardeau limit, the constraint $\xi_T\ll \langle |z| \rangle\/$ simply translates into $k_B T \gg \hbar \omega_z$.
The zero temperature result for the damping is thus only valid in the experimentally hardly
accessible regime $T \ll \hbar \omega_z\/$, while for realistic temperatures the 
damping is expected to increase like $\sqrt{T}\/$. 

\section{Summary}

In summary, we have calculated the damping of the Bogoliubov-Anderson
mode for an attractively interacting 1D Fermi gas in the whole
regime between the BCS and the BEC limit. At zero temperature, the
damping is of a hydrodynamic form with a viscosity $\zeta=\alpha_{\zeta}\hbar n$.
The associated constant $\alpha_\zeta$ is a smooth function along the crossover from a
BCS type superfluid to a BEC of strongly bound pairs of Fermions, with
a universal value $\alpha_\zeta=0.38$ at the unitarity point. It is remarkable, that a
rough analysis \cite{Gelman} of the experiments by Bartenstein et al. \cite{Bartenstein}
gives a value of $0.3$ for the universal viscosity coefficient of the 3D unitary
Fermi gas at the lowest attainable temperatures. However, it is obvious
that a comparison between this and our result in 1D is not meaningful.
Nevertheless, the fact that the Bogoliubov-Anderson mode spectrum and
velocity are hardly different between the one and the three dimensional
case, suggests that the viscosity in 3D exhibits a similar dependence on
the inverse coupling constant $1/\gamma=1/(k_F a)$.
The unitarity point would then define a minimal 
value of the viscosity on the \emph{Fermi} side of the crossover, yet 
lower viscosities are expected by going further into the BEC regime. 
It is a peculiar 
property of the 1D BCS-BEC crossover problem, that the 
boundary between Fermionic and Bosonic behaviour is sharp
and defined by the confinement induced resonance. A similar
sharp separation, however, does not exist in three dimensions.

At finite temperature, the sound damping in 1D 
behaves like $\Gamma_k\sim k^{3/2}\/$ and thus is not 
of a hydrodynamic form. The
resulting damping of the breathing mode in a trapped gas has been
calculated within a simple model, which accounts for the
inhomogeneity in the case of a nonstandard damping.  In particular, it 
has been shown that in the experimentally relevant regime 
$T\gg \hbar\omega_z\/$, the damping increases like $\sqrt{T}\/$, thus
providing a sensitive measure of temperature in
strongly interacting 1D gases. Experimentally, an attractive Fermi gas
near a Feshbach- and confinement induced resonance has been 
realized by Moritz et al. \cite{Moritz:2005}. Since the typical 
temperatures in this gas were of order $T\approx 0.2\, T_F\/$ with
$T_F\approx N\cdot\hbar\omega_z\/$  and typical particle numbers 
are $N\approx 100\/$, the condition $T\gg \hbar\omega_z\/$ is realized.
It would be quite interesting therefore, to study the temperature 
dependence of the breathing mode Q-factor similar to the
measurements performed in 1D Bose gases \cite{Moritz:2003}.  
In this context, it is interesting to note that for the Tonks-Girardeau gas,
exact results for the dynamics have been derived at zero temperature
by  Minguzzi and Gangardt \cite{Minguzzi}. In particular they imply
zero damping of the breathing mode at the unitarity point, i.e. 
an infinite Q-factor. From our present results, the Q-factor is
infinite only in the limit $N\to\infty\/$ but not for the finite and typically
small values $N\approx 50-100\/$ realized experimentally.  This point needs
to be studied further.

\appendix
\section{QHD vs. Bosonization}

As mentioned earlier, the QHD approach does not give rise to interaction terms between spin and charge excitations. We now use Bosonization techniques to construct $H_{int}$ on the BCS side of the crossover, where spin excitations are present, and show that these terms do not contribute to the damping rate.

In order to obtain damping in the Tomonaga-Luttinger model, one needs to to incorporate the quadratic dispersion relation in the Hamiltonian, leading to third order terms in the fields as was already shown by Haldane for a spinless gas of Fermions \cite{Haldane}. Since we are dealing with spin 1/2 particles, third order terms arise which couple charge and spin excitations. Bosonizing the kinetic energy term via point splitting we find
\begin{align*}
H_{int} & = \frac{\hbar^2}{2 m} \sum_{\sigma=\uparrow,\downarrow} \int_0^L dx \Big\{ \partial_x \psi_\sigma^\dagger \partial_x \psi_\sigma + \partial_x \overline\psi_\sigma^{ \dagger} \partial_x \overline\psi_\sigma \Big\} \\ & = -\frac{\hbar^2}{6 m}\sqrt{\frac{\pi}{2}} \int_0^L dx \Big\{ (\partial_x \varphi_c) \Pi_c^2 + \Pi_c^2 (\partial_x \varphi_c) \\ & + \Pi_c (\partial_x \varphi_c)\Pi_c +
 (\partial_x \varphi_c)^3 + 3 \Pi_c \big[(\partial_x \varphi_s)\Pi_s \\ & + \Pi_s(\partial_x \varphi_s)\big]+3(\partial_x \varphi_c)\big[\Pi_s^2+(\partial_x \varphi_s)^2\big] \Big\}
\label{Hint1}
\end{align*}   
where we used the same notation as \cite{Senechal}. A similar result was derived
previously by Fedichev et al. in the context of spin-charge separation in 1D
repulsive Fermi gases \cite{Fedichev}. 
To avoid confusion, it should be mentioned that here $\partial_x \varphi$ plays the role of density fluctuations and $\Pi$ describes phase fluctuations in contrast to (\ref{Hamiltonian}).
The charge-charge-interaction Hamiltonian given above is essentially the same as the one in (\ref{Hamiltonian}) with one subtle difference: in the QHD approach the $\Pi^3$ term has a prefactor $\frac{d}{d\rho_0}(v_c^2/\rho_0)$ which is absent in the bosonized counterpart. This factor is important because it prevents $\alpha_\zeta$ and thus the damping from going to infinity in the BEC limit. Since the crossover from the BCS to the BEC regime is continuous, the damping must also change continuously when one crosses the point of unitarity. This argument leads us to include this prefactor also in $H_{int}$ on the BCS side and thus use (\ref{Hamiltonian}) in the whole crossover regime.

The contribution to the phonon damping rate arising from interaction with spin excitations 
in second order perturbation theory can be calculated from the following diagrams:
\begin{center}
	\begin{fmffile}{fmfdiag4}
	\begin{fmfchar*}(30,25)
  	\fmfleft{i1} \fmfright{o1}
  	\fmf{fermion,label=$k$}{i1,v1}
  	\fmf{fermion}{v2,o1}
  	\fmf{dashes_arrow,label=$k-q$,tension=0.2,left=0.5}{v1,v2}
  	\fmf{dashes_arrow,label=$q$,tension=0.2,left=-0.5}{v1,v2}
  	\fmfdot{v1,v2}
	\end{fmfchar*}
	\end{fmffile}
	
	\begin{fmffile}{fmfdiag5}
	\begin{fmfchar*}(30,25)
  	\fmfleft{i1} \fmfright{o1}
  	\fmf{fermion,label=$k$}{i1,v1}
  	\fmf{fermion}{v2,o1}
  	\fmf{dashes_arrow,label=$q-k$,tension=0.2,right=0.5}{v2,v1}
  	\fmf{dashes_arrow,label=$q$,tension=0.2,left=-0.5}{v1,v2}
  	\fmfdot{v1,v2}
	\end{fmfchar*}
	\end{fmffile}
	
	\begin{fmffile}{fmfdiag6}
	\begin{fmfchar*}(30,25)
  	\fmfleft{i1} \fmfright{o1}
  	\fmf{fermion,label=$k$}{i1,v1}
  	\fmf{fermion}{v2,o1}
  	\fmf{dashes_arrow,label=$-k-q$,tension=0.2,right=0.5}{v2,v1}
  	\fmf{dashes_arrow,label=$q$,tension=0.2,right=-0.5}{v2,v1}
  	\fmfdot{v1,v2}
	\end{fmfchar*}
	\end{fmffile}
\end{center}
For $T=0$ the contribution from the first diagram corresponding to spontaneous decay is given by
\[
\Gamma \sim \int \frac{dq}{2\pi} V_{sc}^2(k,q,k-q) \delta \big( \underbrace{v_c k - \omega_s(q)- \omega_s(k-q)}_{\neq 0 \; \forall q}\big)=0
\]
where $V_{sc}$ denotes the spin-charge interaction vertex and $\omega_s(q)$ is the spinon dispersion relation. We immediately see that this process is forbidden by energy conservation. The other two diagrams give a small contribution only for $T>0$. In the strong coupling limit ($1/\gamma \rightarrow 0^{-}$) we obtain
\[
\Gamma_k^{sc} \approx \frac{1}{64 \pi^3} \frac{\varepsilon_F}{\hbar} \; \gamma^4 \; \xi_T k \; \text{e}^{-\beta \Delta/2}+ \mathcal{O} \big( (\xi_T k)^2 \big) 
\]
This term involves the energy gap $ \Delta \sim 2 \varepsilon_F \gamma^2/\pi^2 $ and thus is negligible compared to (\ref{dampT}).



\begin{thebibliography}{99}
\bibitem{sqrt2}Note the $\sqrt{2}$ difference in our definition 
of $a_{\perp}$ compared with 
\cite{Olshanii,BMO}, which accounts for the difference in the 
value of $A\equiv 1.0326$.
\bibitem{charge} The notation $v_c\/$ for the velocity of the Bogoliubov-Anderson
mode reflects the usual notation in the Luttinger liquid context as a 
'charge' velocity, yet in the present case of neutral atoms, it is a zero sound type
density oscillation in a superfluid system.
\end{thebibliography}
\end{document}